\documentstyle[12pt,aasms4,flushrt]{article}
\def\ang{\thinspace{\rm \AA}}
\def\civ{{\rm C}\thinspace{\sc{iv}}}
\def\eg{{\it e.g.}}
\def\erg{${\rm ergs\ cm}^2\ {\rm s}^{-1}$}
\def\etal{{\it et~al.\/}}
\def\ha{\ifmmode {{\rm H}\alpha}
        \else {H$\alpha$}\fi}
\def\hnought{\ifmmode H_0
    \else $H_0$\fi}

\def\kms{~{\rm km\ s}^{-1}}
\def\la{\ifmmode {{\rm Ly}\alpha}
        \else {Ly$\alpha$}\fi}
\def\lstar{{$L_{\ast}$}}
\def\msun{{$M_{\odot}$}}
\def\oii{[O\thinspace{\sc{ii}}]}
\def\oiii{[O\thinspace{\sc{iii}}]}
\def\p.{^{\prime\prime}\kern-2.1mm .\kern+.6mm}
\def\qnought{\ifmmode q_0
    \else $q_0$\fi}
\def\s.{^{s}\kern-2.1mm .\kern+.6mm}
\def\ten#1{\ifmmode 10^{#1}
    \else $10^{#1}$\fi}
\slugcomment{}

\begin{document}

\title{Detection of a Lyman {$\boldmath \alpha$} Emission-Line Companion to\\
  the {$\boldmath z$} = 4.69 QSO BR1202--0725\altaffilmark{1}}

\author{Esther M. Hu}
\affil{Institute for Astronomy, University of Hawaii, 2680 Woodlawn Dr.,
  Honolulu, HI 96822\\
  hu@ifa.hawaii.edu}
\author{Richard G. McMahon}
\affil{Institute of Astronomy, University of Cambridge, Madingley Road,
  Cambridge CB3\thinspace{0HA}\\
  rgm@ast.cam.ac.uk}
\and
\author{Eiichi Egami\altaffilmark{2}}
\affil{Institute for Astronomy, University of Hawaii, 2680 Woodlawn Dr.,
  Honolulu, HI 96822\\
  egami@mpe-garching.mpg.de}
\altaffiltext{1}{Based on observations with the NASA/ESA {\it Hubble Space
  Telescope} obtained at the Space Telescope Science Institute, which is
  operated by AURA, Inc., under NASA contract.}
\altaffiltext{2}{Current address: Max-Planck-Institut f\"ur extraterrestrische
  Physik, Postfach 1603, 85740 Garching bei M\"unchen, Germany}

\begin{abstract}
We report the detection of a nearby emission line companion to the $z=4.695$
quasar BR1202--0725.  Deep narrow-band exposures on this field from the UH
2.2~m show a \la\ flux of $1.5\times\ten{-16}$ ergs cm$^{-2}$ s$^{-1}$.
High-resolution $HST$ WFC2 imaging in the F814W filter band shows continuum
structure near the emission position, at $2\p.6$ NW of the quasar,
corresponding to a projected separation of \hbox{$\sim7.5\,h^{-1}$} kpc for
\qnought=0.5, where $h\equiv \hnought/100\kms$ Mpc$^{-1}$.  We discuss possible
explanations for the combined line and color properties.  The ionization
is most likely produced by the quasar, but if due to underlying star
formation would require a star formation rate of $\sim7\,h^2$ \msun\
yr$^{-1}$.
\end{abstract}


\section{Introduction}
The known high-$z$ quasars are extremely luminous objects capable of
lighting up any gas or dust in their vicinity out to several hundred
kiloparsecs by direct ionization, and by scattering in \la\ and continuum
light.  At redshifts $z\sim4$, when the age of the universe was $\sim1$
billion years, the available time is comparable with the dynamical
timescales of galactic halos and any infalling gas within a quasar host
galaxy should be illuminated by the central source (\cite{ree88}).  A
number of \la\ searches around high-redshift quasars have succeeded in
identifying \la\ emitters at the redshift of the quasar and within a
small angular separation from the quasar (\eg, \cite{djorg85,hu91}).  The
simplest explanation for the {\la}-emitting gas seen around these systems
is that it is due to ionization of the gas in nearby, possibly
interacting, galaxies or gas clouds by the quasar (\cite{hu87,hu91}).
Even very small column densities of gas or dust (\eg, as little as
$10^{18}$ cm$^{-2}$ in neutral hydrogen) can result in significant
scattering or reprocessing.

In the present Letter we describe the discovery of a very high redshift
emission companion to the $z=4.695$ quasar BR1202--0725, designated here
as BR1202--0725$e$, which is detected in deep narrow-band images centered
on the quasar's \la\ emission, at a position $\sim2\p.3$ northwest of the
quasar.  At $z=4.69$, $1''$ corresponds to $3.0\,h^{-1}$ kpc for
\qnought=0.5 (or $5.2\,h^{-1}$ kpc for \qnought=0.1), so $2\p.3$
corresponds to a projected separation of $6.9\,h^{-1}$ ($12.0\,h^{-1}$)
kpc.  The object is also seen in an $HST$ $I$ band continuum exposure
(F814W), with a centroid slightly offset from the emission, and has been
studied in several colors by D'Odorico \etal\ \markcite{dodor95}(1995),
who argued based on the optical colors, with additional evidence from
near-IR magnitude measurements by  Djorgovski \markcite{djorg95}(1995),
that this object is at $z>4$, but identify it with the $z=4.38$ damped
\la\ system discovered by Storrie-Lombardi \etal\ \markcite{lsl95}(1995)
rather than with the quasar itself.  The present data suggest that the
continuum object is at the redshift of the quasar rather than at the
redshift of the foreground damped \la\ system.

\section{Data}

The QSO BR1202--0725 was discovered in the APM BRI survey for $z>4$ QSOs
(\cite{apm}).  As part of a program to search for high-$z$ objects in the
fields of $z>4$ quasars, we have obtained a number of exposures through a
narrow-band filter centered on the quasar's redshifted \la\ emission
(central wavelength 6925\ang, 80\ang\ bandpass) and through $B$, $I$, $K$
multi-color imaging with a Tektronix 2048$^2$ camera at optical
wavelengths and the NICMOS3 ($256^2$) and QUIRC ($1024^2$) cameras at
near-infrared wavelengths\footnote{The IR observations cited here are
taken using the $K'$ filter, which has a central wavelength of 2.1
$\mu$m, in order to suppress the thermal component of background.  We
will refer to $K$ and $K'$ interchangeably here; the detailed photometric
conversion is given in Wainscoat \& Cowie \markcite{wai92}(1992).}.  The
optical data were taken as a series of sky noise-limited integrations,
each with an exposure time of 30 minutes (in the case of the narrow-band
exposures), with an offset step of $10''$ between successive frames, and
a median sky flat was generated from the on-field exposures for each
night, while the IR data followed standard deep IR imaging procedures
(\eg, \cite{ksurvey_1}).  The optical data were taken on the UH 2.2 m on
the nights of (UT) 5--7 March 1994, 7--8 April 1994, and 29--30 March
1995, while the IR data were taken on 16--18 April 1994 (NICMOS3) and on
18--20 March 1995 (QUIRC) using the UH 2.2 m.  The total exposure time
was 8 hrs in \la, 4.2 hrs in $B$, 5.25 hrs in $I$, 1.5 hrs in $I'$
(8340/895), and 17 hrs in $K'$.  The broad-band data were calibrated
using Landolt standards (\cite{lan92}) and stars drawn from both the
UKIRT faint standards (\cite{ukirt}) and  Elias standards (\cite{elias})
in observations both before and after the target exposures.  The
narrow-band exposures were calibrated using Feige 34 and BD+33 2642.  In
addition, a relative calibration check with Landolt standard-calibrated
continuum exposures taken in an 895\ang\ line-free band centered at
8340\ang\ was also used to verify estimates of the \la\ and continuum
flux.

A series of $HST$ exposures were obtained on 25--26 July 1995 using
WFPC2 and the F814W filter. The F814W filter has an effective central
wavelength of $\sim$7900\ang\ and a effective width of $\sim$1450\ang,
sampling the spectral range 1260\ang\ to 1514\ang\ in the rest frame of
the QSO (a region free of strong lines such as \la\ and \civ).  $HST$
exposures were taken as a sequence of eight exposures, with the two
initial exposures each being 1000 s long and the subsequent six exposures
each 1200 s in duration, over four primary orbits for a total of 9200 s.
Object magnitudes on the $HST$ F814W data were obtained using the
PHOTFLAM calibrated fluxes over a $2''$ diameter aperture and converted
to a Kron-Cousins $I$ band magnitude following Cowie, Hu, \& Songaila
\markcite{cowie95}(1995).

The \la\ image of the quasar shows a substantial extension to the
northwest of the quasar and centered at a radial separation of $2\p.3$.
In Figure 1 (left panel) we show a sharpened version of this image which
was deconvolved using maximum entropy from typical FWHM of $\sim0\p.9$ on
individual exposures to a FWHM of $0\p.6$ with the point spread function
taken from the star which can be seen in the upper north west corner of
the image.  The $HST$ $I$-band image shows a linear galaxy, extended
towards the quasar, also lying at this position, though on the westward
side of the \la\ emission (Fig.~1; right panel).  This slight
displacement ($\sim0\p.6$ in the east-west direction) appears to be
real.  The continuum object agrees in position with the object detected
by D'Odorico \etal\ \markcite{dodor95}(1995) using the ESO NTT.
Subtracting the continuum we find a \la\ flux of $1.5\times\ten{-16}$
\erg\ for the companion, and an $I$ (Kron-Cousins) = $24.2\pm0.1$ for the
associated continuum object.  The emission is comparable to the measured
fluxes found for other quasar \la\ companions (\eg, \cite{djorg87,hu91}),
which are typically a few times $\ten{-16}$ \erg, and the continuum
magnitude is in good agreement with D'Odorico \etal\ \markcite{dodor95}%
(1995), who find $B, V, R, I$= ($>27.3$, 26.5, 24.3, 24.1).  Our $K'$
band observations shown in Fig.~2 also show an extension to the northwest
of the quasar.  (There are hints in both the $HST$ and $K'$ data of
further emission extending in a line to larger radii.)\ \ Measuring in a
$1''$ diameter aperture centered on the $I$ band object we find
$K=23.4\pm0.3$ when the underlying quasar contribution is subtracted,
while Lu \etal\ \markcite{lu96}(1996) quote a $K$ magnitude of $23\pm1$
based on Djorgovski's \markcite{djorg95}(1995) observations.  D'Odorico
\etal\ point out that the sharp break in the optical implies that the
object is at high redshift ($z > 4$), consistent with its identification
as a quasar companion.  The combined SED is illustrated in Fig.~3 where
it is compared with the spectrum of the quasar itself, showing that the
photometry on the companion is consistent with its lying at the quasar
redshift.  The association of the continuum object with the quasar
\la\ strongly suggests that it is {\em not\/} associated with the damped
\la\ system at $z=4.38$ as D'Odorico \etal\ and Lu
\etal\ \markcite{lu96}(1996) have recently suggested.  A second galaxy
lying about $3 {1\over2}''$ to the SW of the quasar with a slightly
fainter $I$ magnitude (24.5) may represent an alternative candidate,
since we do not see any strong \la\ emission at the quasar redshift
associated with this object.

The present results indicate that some caution must be applied to
identifying objects in close proximity to high-redshift quasars with
foreground absorption systems.  Steidel, Sargent, \& Dickinson
\markcite{stei91}(1991) and Arag\'{o}n-Salamanca \markcite{ara95}(1995)
have serendipitously discovered \la-emitting companions to $z\sim3$ QSOs
when attempting to identify foreground absorbers.  At high redshift the
difference in distance modulus between the background QSO and an
intervening absorber is much less than for a typical case at low
redshift, so that any faint object lying in the vicinity of a QSO may be
as likely to be associated with the QSO as with the absorption-line
system.

The discovery of a \la\ companion near BR1202--0725 indicates that it
is fruitful to search for such objects at optical wavelengths.  More
precise information on the nature of this object will require
spectroscopic data at IR and optical wavelengths, which will also
provide physical diagnostics on the emission system, to distinguish
between AGN-like or quasar-excitation mechanisms and to supply additional
information on the nature of this system.

\section{Conclusions}

Since the rest-frame $B$ magnitude is roughly coincident with the
observed $K$ magnitude we may directly obtain the rest-frame absolute
$B$ magnitude ($M_B$) as
\begin{equation}
  {M_B = m_K - dm +4.1}
\end{equation}
where $dm$ is the distance modulus and the last term is the K correction.
For \qnought=0.5 and \hnought=$50\kms$ Mpc$^{-1}$ we find $M_B=-20.5$ or
only about $1/3$ of the local \lstar\ (\cite{loveday}), so the galaxy
is roughly comparable to an \lstar\ galaxy.  As can be seen from Fig.~3 its
SED is similar to that of the quasar but it contains only about 0.1\% of
the quasar light.

The observed equivalent width of the \la\ is about 450\AA, corresponding
to a rest-frame value of 80 \AA.  This would be consistent with
excitation by the underlying population (\cite{charl93}) but given the
apparently more extended nature of the ionized gas, its slight
displacement from the continuum centroid, and its proximity to the much
more luminous quasar, it is more probable that the quasar is the primary
ionizing source.  The ionized gas covers a sufficiently large fraction of
the surface surrounding the quasar so that only a very small fraction of
the ionizing flux passing through it needs to be absorbed and reradiated
to produce the observed \la\ emission (\cite{hu87}).  However, Pahre \&
Djorgovski \markcite{pahre95}(1995), who have performed targeted IR
narrow-band searches around BR1202--0725 in the \oii\ line using NIRC at
Keck (with three other fields studied in \ha\ or \oiii), saw no evidence
for extended emission near the quasar down to fluxes of
$1.60\times\ten{-17}$ \erg, giving \oii/\ha$\sim0.1$.  This would require
that there is very little extinction of the \la\ photons if metallicities
were near solar, but it is likely that the metallicities are much lower
in these early objects.  The proximity of the quasar (separation
$\sim2''$ with typical seeing FWHM $\sim0\p.75$ in the Pahre \&
Djorgovski \markcite{pahre95} observations) may also imply that somewhat
higher flux limits on \oii\ (\eg, by roughly a factor of two) are
appropriate at the position of BR1202--0725$e$.  If the observed emission
were due to photoionization by stars, with little attenuation by dust,
then the luminosity in the line of $7\times\ten{42}\,h^{-2}$ ergs
(\qnought=0.5) corresponds to a star formation rate (SFR) of $\sim7\,h^{-2}$
\msun\ yr$^{-1}$, where we use Kennicutt's \markcite{kenn83}(1983)
relation between \ha\ luminosity and
SFR=$L($\ha)$\times8.9\times\ten{-42}$ ergs s$^{-1}$ \msun\ yr$^{-1}$,
and assume \la/\ha=8.7 for Case B recombination (\eg, \cite{brock}).

Close \la\ companions to the high-$z$ quasars are only infrequently
seen at these fluxes (about 15\% of the cases in \cite{hu91}, all of
which were around radio-loud quasars, where they were detected in
roughly a third of the cases, and generally at lower flux levels) and
this is also true of the $z >4$ quasars, where BR1202--0725 is the only
one of five radio-quiet cases where we see such emission.  They may
represent cases where a neighboring gas cloud or galaxy (as might be the
case here) is interacting and merging with the underlying quasar host
producing enough extended gas to form a significant \la\ companion when
ionized by the quasar.

Finally, it is notable that BR1202--0725 has been detected at mm
wavelengths (\cite{rgm94,submm94}).  This radiation is consistent with
thermal emission from \ten{9} \msun\ dust analogous to that detected by
$IRAS$ in nearby star-forming galaxies, and this would also be consistent
with dust originating in an interaction with the quasar.

\acknowledgments
This research was partially supported by STScI grant GO-5975 and in part by
the University of Hawaii.  E. M. H. would also like to gratefully acknowledge
a University Research Council Seed Money grant.  R. G. M. acknowledges the
support of the Royal Society.

\newpage

\begin{figure}[h]
\caption{The $25\p.6\times25\p.6$ field surrounding the $z=4.69$ quasar,
BR1202--0725 at coordinate center [$\alpha(1950)\!:
12^h\,02^m\,49\s.19,\ \delta(1950)\!: -7^{\circ}\,25'\,50\p.4$; McMahon
\etal\ {\protect{\markcite{rgm94}}}(1994)] (left panel) seen in an 8 hr
exposure on
the UH 2.2-m telescope through an 80\ang\ wide, narrow-band filter
centered on the quasar's redshifted Ly$\alpha$ emission, and in a 2.6 hr HST
exposure (right panel) through the F814W (`wide $I$') continuum band.
The narrow-band image has been sharpened using maximum entropy methods to
highlight the position of the faint companion emission against the bright
quasar signal, and this feature is shown encircled by a 2$''$ diameter
aperture.  The enclosed emission line flux is $\sim1.5\times 10^{-16}$
ergs cm$^{-2}$ s$^{-1}$ --- comparable with the Ly$\alpha$ brightness of
other close companions to high-redshift quasars (\eg,
{\protect{\cite{djorg87,hu91}}}).  The faint continuum feature seen near the
position
of the Ly$\alpha$ emission is located $1\p.63$ west and $2\p.08$ north of the
quasar, and has a measured $I$ magnitude of 24.2. By contrast, the
Ly$\alpha$ emission lies $1\p.0$ west and $2\p.1$ north of the quasar. North is
up and East is to the left.}
\end{figure}
\begin{figure}[h]
\caption{Comparison of $K$ band image (left panel) of the region near the
quasar BR1202--0725 with the $HST$ F814W (`wide-$I$', right panel) image,
showing the extension in continuum light to the northwest in both
images.  Each field is $22''\times22''$ and the plate scale matches that
of Fig.~1.  The $K$-band image has been sharpened with maximum entropy
using the star in the upper right of the frame in order to highlight the
faint structure near the quasar.  The composite $K$ band image represents
a 17 hr integration at the UH 2.2 m, and the estimated $K$ magnitude
measured in a 1 arcsec diameter aperture centered on the position of the
$HST$ $I$-band continuum source is $23.4\pm0.4$ after subtracting the
underlying quasar contribution.}
\end{figure}
\begin{figure}[h]
\caption{The spectral energy distribution of the quasar (filled squares)
and its \la\ companion (horizontal bars) as measured at $B$, $I$, and
$K'$, and in the 6925\ang\ narrow-band filter.  The observed wavelength
from 3000\ang\ to 3$\,\mu$m is shown plotted logarithmically;
corresponding rest-frame wavelengths (from 527\ang\ to 5268\ang) are
indicated by the tick marks along the top axis.  The width of each bar
reflects the filter bandpass.  The $B$, $I$, and $K$ photometry for the
quasar is taken from Egami {\protect{\markcite{egami}}}(1995), and these
imaging
data sets have been combined with the narrow-band imaging data for the
measurements shown here. The $I$ data for the \la\ companion is taken
from the 2.6 hr $HST$ WFPC2 exposure, and the \la\ narrow-band and $K$
images have been sharpened with maximum entropy techniques as shown in
Figs.~1 and 2 to allow separation from, and subtraction of the underlying
quasar light in these bands.  We have supplemented these measurements for
the companion with the upper limit on $B$ and the $V$ magnitude given by
D'Odorico \etal\ {\protect{\markcite{dodor95}}}(1995); their $R\/$ band is
heavily
contaminated by emission, and is not used here.  One sigma error bars are
shown.}
\end{figure}
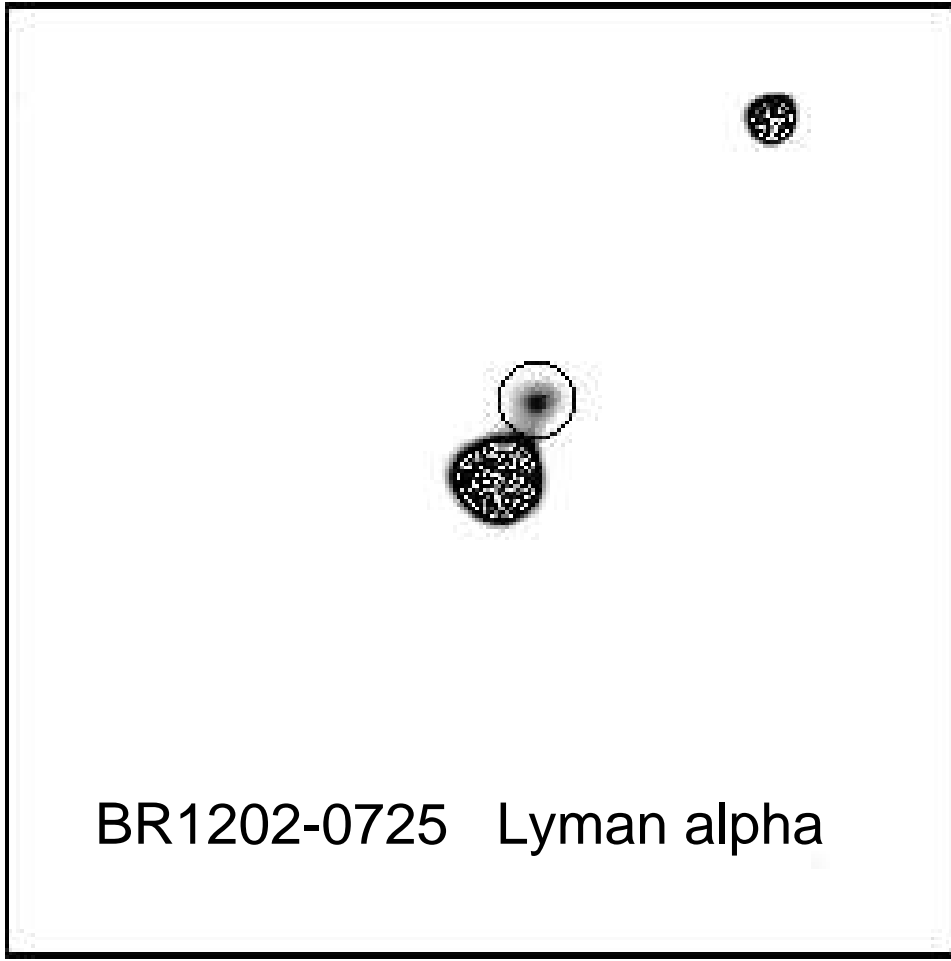
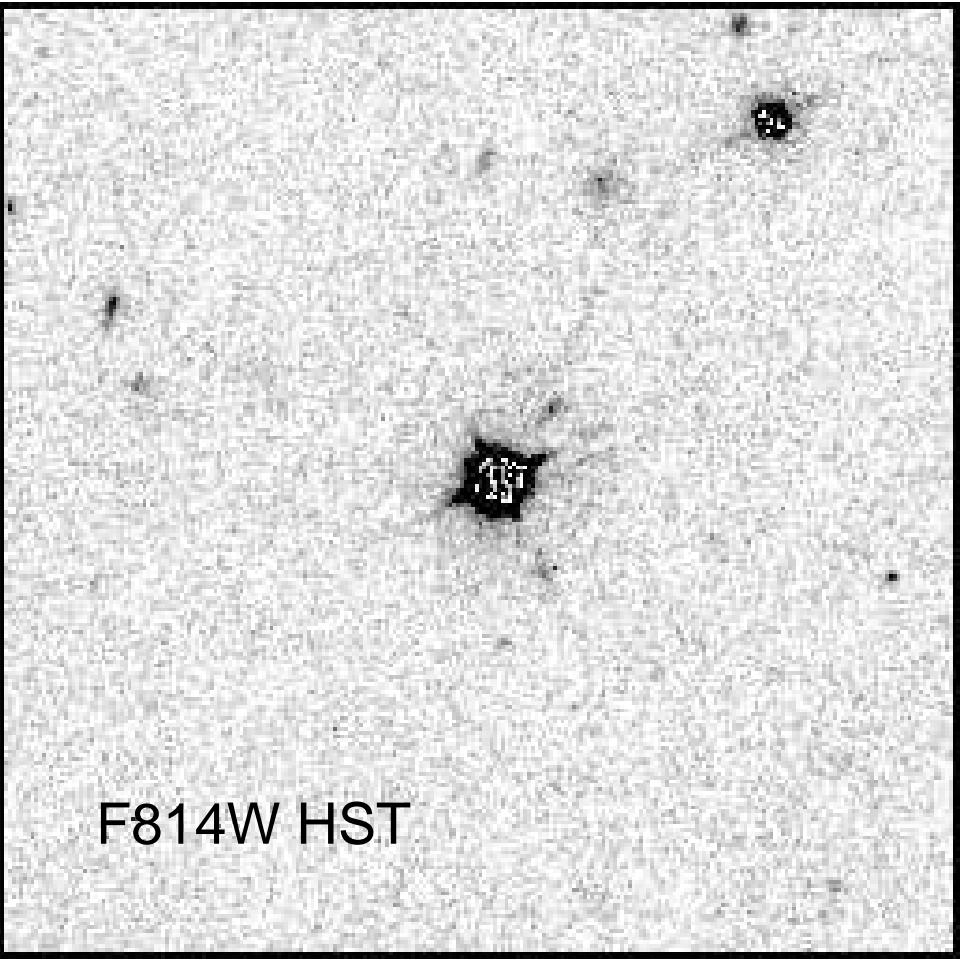
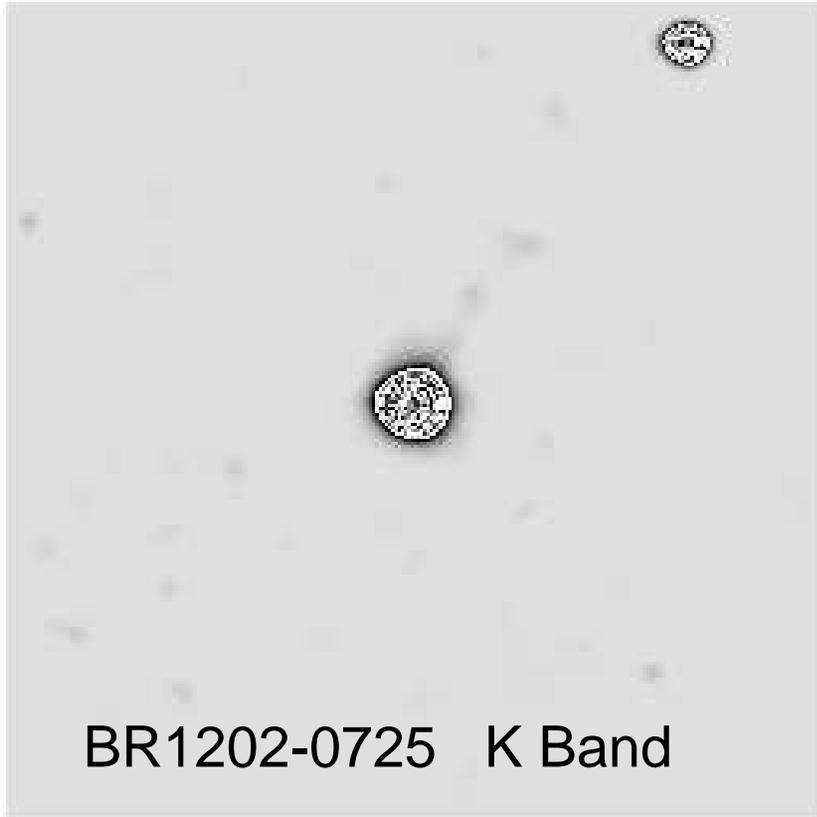
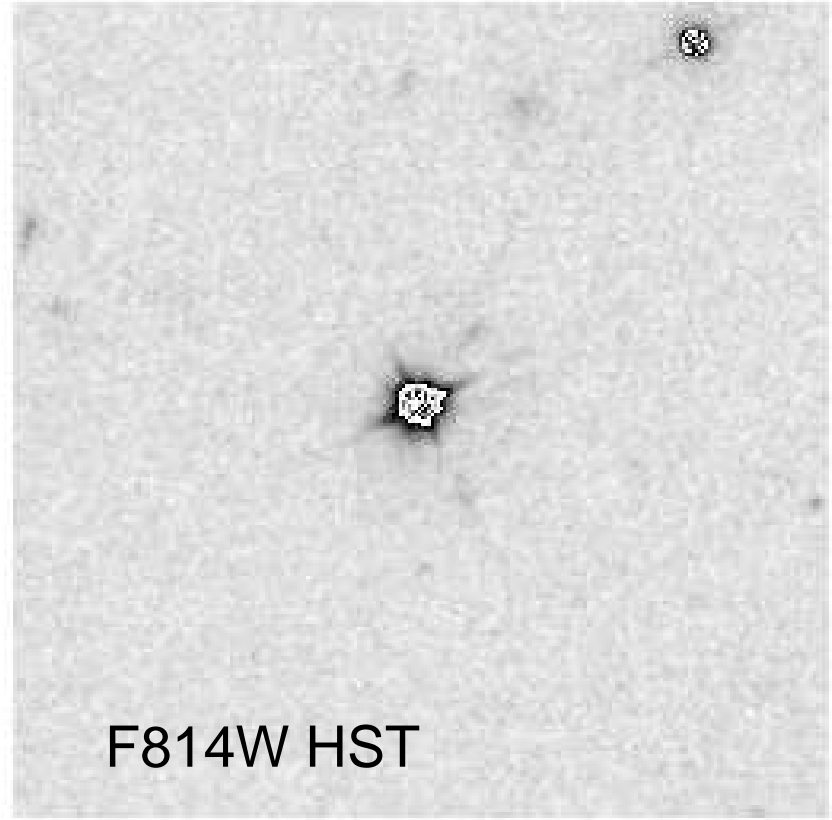
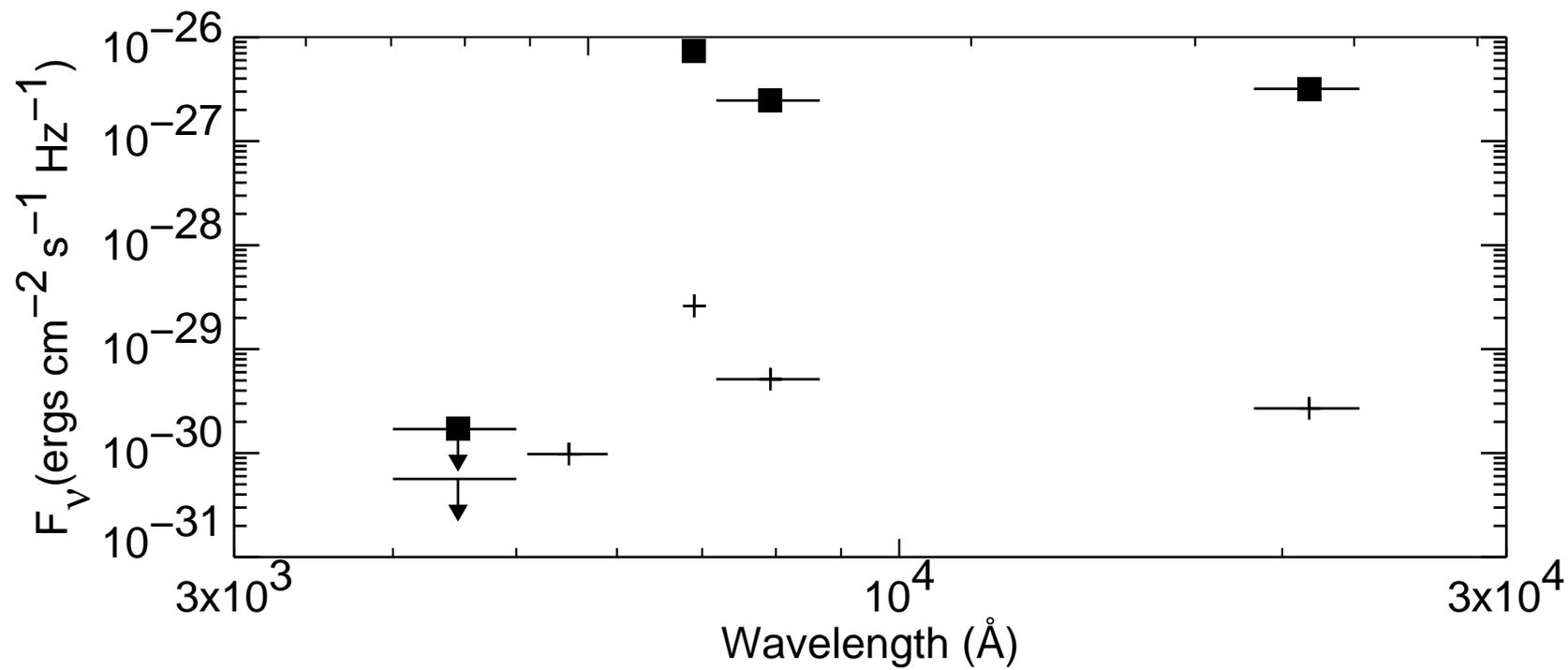
\end{document}